\documentclass[conference]{IEEEtran}
\IEEEoverridecommandlockouts
\usepackage{tabularx}  
\usepackage{cite}
\usepackage{amsmath,amssymb,amsfonts}
\usepackage{algorithmic}
\usepackage{graphicx}
\usepackage{textcomp}
\usepackage{xcolor}
\usepackage{booktabs} 
\def\BibTeX{{\rm B\kern-.05em{\sc i\kern-.025em b}\kern-.08em
    T\kern-.1667em\lower.7ex\hbox{E}\kern-.125emX}}
\begin{document}

\title{NeuroHD-RA: Neural-distilled Hyperdimensional Model with Rhythm Alignment\\
}
\author{
\IEEEauthorblockN{ZhengXiao He, Jinghao Wen, Huayu Li, Siyuan Tian, and Ao Li*}
\IEEEauthorblockA{
ZhengXiao He, Huayu Li, and Ao Li are with the Dept. of Electrical and Computer Engineering, \\
University of Arizona, Tucson, AZ, USA\\
Jinghao Wen is with the Dept. of Electrical and Computer Engineering, 
Villanova University, Villanova, PA, USA\\
Siyuan Tian is with Microsoft Research, Shanghai\\
Email: \{zhengxiaohe, hl459, aoli1\}@arizona.edu, jwen01@villanova.edu, siyuan.tian@microsoft.com\\
*Corresponding author: Ao Li
}
}

\maketitle

\begin{abstract}
We present a novel and interpretable framework for electrocardiogram (ECG)-based disease detection that combines hyperdimensional computing (HDC) with learnable neural encoding. Unlike conventional HDC approaches that rely on static, random projections, our method introduces a rhythm-aware and trainable encoding pipeline based on RR intervals, a physiological signal segmentation strategy that aligns with cardiac cycles. The core of our design is a neural-distilled HDC architecture, featuring a learnable RR-block encoder and a BinaryLinear hyperdimensional projection layer, optimized jointly with cross-entropy and proxy-based metric loss. This hybrid framework preserves the symbolic interpretability of HDC while enabling task-adaptive representation learning. Experiments on Apnea-ECG and PTB-XL demonstrate that our model significantly outperforms traditional HDC and classical ML baselines, achieving 73.09\% precision and an F1 score of 0.626 on Apnea-ECG, with comparable robustness on PTB-XL. Our framework offers an efficient and scalable solution for edge-compatible ECG classification, with strong potential for interpretable and personalized health monitoring.
\end{abstract}

\begin{IEEEkeywords}
Electrocardiogram (ECG), Hyperdimensional Computing (HDC), Sleep Apnea Detection, Metric Learning
\end{IEEEkeywords}

\section{Introduction}
The increasing availability of large-scale physiological datasets has enabled significant advances in machine learning for health monitoring. In particular, electrocardiogram (ECG) signals provide critical insights into cardiopulmonary conditions such as sleep apnea \cite{penzel2000apnea,silva2009longitudinal}, as well as a wide range of cardiac disorders, exemplified by the PTB-XL diagnostic ECG database \cite{wagner2020ptb}. Deep neural networks (DNNs) have demonstrated strong performance in these tasks \cite{rajpurkar2017cardiologist,ismail2020inceptiontime,wang2024medformer}. However, their computational complexity and limited suitability for edge deployment hinder their adoption in real-world clinical settings, particularly in resource-constrained environments.
Importantly, many clinical applications do not require overly complex black-box models with extensive functionalities. In practice, hospitals often prioritize models that are lightweight, efficient, and easy to deploy—particularly for routine screening, risk stratification, or integration into existing workflows. These practical constraints motivate the development of streamlined learning paradigms that emphasize efficiency and scalability, rather than maximizing model complexity.

Hyperdimensional computing (HDC), inspired by brain-like distributed representation, has emerged as a promising framework for fast, noise-tolerant, and energy-efficient learning \cite{kanerva2009hyperdimensional}. In HDC, inputs are encoded in high-dimensional binary or bipolar vectors and classified according to their similarity to class prototypes stored in associative memory, using operations such as binding, bundling, and permutation \cite{rahimi2016robust}. Owing to its bitwise parallelism, low arithmetic complexity, and robustness to noise, HDC is particularly well-suited for edge computing, and its model parameters can be stored in bitwise form instead of floating-point, greatly reducing memory and energy consumption.

However, classical HDC systems are constrained by limited adaptability. Their encoders are typically handcrafted and task-specific, and class hypervectors are constructed using non-trainable aggregation techniques such as averaging or majority voting. As a result, their ability to model complex or heterogeneous physiological signals remains limited.
To address these limitations, recent research has introduced Learnable HDC (LeHDC) architectures that integrate symbolic HDC inference with trainable neural components \cite{duan2022lehdc,imani2017voicehd}. However, these efforts have mainly focused on synthetic or low-complexity datasets such as MNIST \cite{lecun1998mnist}, Fashion-MNIST \cite{xiao2017fashion}, and CIFAR-10 \cite{krizhevsky2009learning}, limiting their applicability to real-world biomedical tasks.

To address these limitations, we propose a neural-to-HDC distillation framework, in which a shallow neural network is trained end-to-end to learn both the input encoder and class-level hypervectors. After training, the learned parameters are reinterpreted as symbolic memories, serving as a high-dimensional encoder and associative memory, thus enabling efficient inference through symbolic reasoning.

Our framework outperforms traditional HDC and classical ML baselines while remaining significantly lighter and more interpretable than deep neural networks, offering a promising foundation for real-world, edge-compatible ECG monitoring. There are three key innovations introduced  in the framework to enhance the physiological relevance and discriminative power of HDC:

\textbf{\textit{Rhythm-aligned encoding}}: We segment ECG signals based on RR intervals to construct cycle-aware representations that preserve heart rate variability and temporal dynamics. While RR-based segmentation has been explored in convolutional networks, our work is the first to integrate RR-guided temporal alignment into the HDC framework. Each RR segment is encoded into a high-dimensional vector, and sequence-level information is preserved through symbolic binding and aggregation, introducing a rhythm-aware symbolic encoding pipeline.

\textbf{\textit{Discriminative hypervector learning}}: We incorporate proxy-based metric learning to enhance the separability of HDC embeddings. Unlike conventional HDC methods that average training vectors to form class prototypes, we learn proxy hypervectors for each class and train the model using a triplet-style loss, encouraging intra-class compactness and inter-class margins. To our knowledge, this is the first use of proxy-based metric learning in an HDC setting for physiological signal modeling, enabling a more structured and discriminative high-dimensional space.

\textbf{\textit{Neural-symbolic encoder for learnable high-dimensional projection}}: Instead of relying on fixed random encoders, we introduce a shallow and trainable neural encoder that maps RR-aligned ECG segments to symbolic high-dimensional vectors. This encoder is optimized end-to-end and binarized after training to form task-adaptive item memories, bridging neural representation learning with symbolic inference. This design improves the specificity of the encoding and the interpretability of the model while retaining the lightweight nature of HDC.

We evaluate our framework on two representative ECG classification tasks:
Apnea-ECG: a sequence-level sleep apnea detection task \cite{penzel2000apnea}, and
PTB-XL: a large-scale diagnostic benchmark with 12-lead ECG waveforms \cite{wagner2020ptb}. Our experiments demonstrate that the proposed model achieves competitive classification performance while preserving the efficiency and generalizability of symbolic HDC inference. This hybrid paradigm supports flexible training and low complexity deployment, offering a promising solution for reliable and edge-compatible physiological monitoring.

\section{Related Work}

\subsection{HDC in Biomedical Applications}

HDC has recently gained interest in biomedical applications due to its robustness and efficiency under noisy and heterogeneous data conditions. It has been used for medical imaging \cite{ma2022hdcoin}, physiological signal classification (e.g., EMG, EEG, EOG) \cite{rahimi2018efficient, moin2018emg}, and neural activity modeling \cite{ni2024hyperdimensional}.
More recently, efforts have been made to enhance the expressiveness and flexibility of HDC deployment in biomedicine. For example, Ma et al. \cite{ma2024hyperdimensional} proposed a distillation-based framework that transfers knowledge from neural networks to HDC models, improving both the efficiency of inference and the representational capacity. In the medical domain, Chen et al. developed lightweight HDC models for the detection of sleep apnea using ECG and PPG signals~\cite{chen2024lightweight,chen2024energy}, demonstrating the feasibility of symbolic computation in physiological monitoring. However, many of these approaches still rely on fixed encoding schemes that lack flexibility and discriminative power for more complex tasks.
Moreover, recent distillation-based HDC frameworks, while improving inference accuracy, often compromise core HDC principles such as symbolic transparency and bitwise composability, making them less suitable for interpretable or logic-driven applications.

\subsection{Metric Learning with Proxies and Triplets}

Metric learning approaches such as triplet loss and proxy-based methods have shown effectiveness in structuring embedding spaces with well-separated classes \cite{schroff2015facenet, movshovitz2017no}. Proxy-based learning, in particular, stabilizes optimization and scales well with large datasets. To our knowledge, this strategy has not been explored in HDC for physiological signals. In this work, we introduce proxy-based metric learning into the HDC framework, enabling end-to-end optimization of both input encoders and class hypervectors.

\subsection{Lightweight Models for ECG Classification}

While deep learning models such as CNNs \cite{rajpurkar2017cardiologist}, transformers \cite{wang2024medformer}, and attention-based architectures \cite{ismail2020inceptiontime} have achieved strong ECG classification performance, they often require substantial computation and memory. Lightweight HDC-based models have been proposed for ECG and PPG-based sleep apnea detection \cite{chen2024lightweight, chen2024energy}, but they rely on non-learnable encoders and do not incorporate margin-based metric learning. Our method fills this gap by introducing a trainable, rhythm-aware, and proxy-enhanced HDC framework for efficient ECG-based disease classification.

\section{Proposed Method}
\subsection{Background: Hyperdimensional Computing}

Hyperdimensional computing (HDC) encodes data using high-dimensional binary or bipolar vectors, called hypervectors, typically with dimensionality $D \gg 1{,}000$. These hypervectors are combined using symbolic operations such as:

\begin{itemize}
    \item \textbf{Binding} — associative pairing:
    \begin{equation}
        \mathbf{z} = \mathbf{x} \odot \mathbf{y}
    \end{equation}
where $\odot$ denotes element-wise multiplication.

    \item \textbf{Bundling} — aggregation of multiple hypervectors:
    \begin{equation}
        \mathbf{z} = \text{sign}\left( \sum_{i=1}^{N} \mathbf{x}_i \right)
    \end{equation}
\end{itemize}

To handle real-valued physiological signals (e.g., ECG), HDC often first applies a quantization step. A local segment or feature (e.g., mean amplitude in a window) is quantized into a finite number of levels. For instance, a 16-level scheme would discretize input values into a set $\mathcal{B} = \{b_1, b_2, \dots, b_{16}\}$.
Each discrete level $b_k$ is then mapped to a randomly generated high-dimensional vector via an item memory:
\begin{equation}
    \mathcal{M}(b_k) = \mathbf{h}_k \in \{-1, +1\}^D
\end{equation}
This allows a continuous input signal to be symbolically encoded. For example, if a windowed ECG segment is assigned to bin 9, this bin ID can be directly mapped to $\mathbf{h}_9$, which represents that quantized state in the high-dimensional space.
Sequences of such symbols can be bundled into a global representation:
\begin{equation}
    \mathbf{X} = \text{sign} \left( \sum_{t=1}^{T} \rho^t\left( \mathcal{M}(b_t) \right) \right)
\end{equation}
where $T$ is the number of segments or observations, $\mathcal{M}(b_t)$ is the item memory vector for bin $b_t$, and $\rho^t(\cdot)$ denotes a $t$-step cyclic permutation applied to encode temporal order.
This positional encoding is especially important when the input signal is divided into fixed-length blocks rather than physiologically meaningful cycles (e.g., R-R intervals). In contrast, when using rhythm-aligned segmentation, such as extracting one symbol per R-R interval, the sequence implicitly encodes temporal structure, and permutation may be omitted. Thus, HDC allows flexible design choices depending on whether alignment or positional encoding is preferred.

At inference time, the bundled vector $\mathbf{X}$ is compared against class prototypes $\mathbf{c}_j$ using cosine similarity:
\begin{equation}
    \hat{y} = \arg\max_j \frac{\mathbf{X} \cdot \mathbf{c}_j}{\|\mathbf{X}\| \|\mathbf{c}_j\|}
\end{equation}
The symbolic encoding process enables efficient and robust representation of biomedical signals, especially in scenarios with limited resources or real-time constraints.
\subsection{Motivation for Neural-Distilled HDC}

While traditional HDC frameworks are lightweight and robust, they suffer from limited expressiveness due to fixed encoders and non-learnable class representations. Prior studies~\cite{ma2024hyperdimensional} have explored hybrid architectures that incorporate shallow neural encoders into HDC pipelines, showing improved performance on toy benchmarks such as MNIST and MedMNIST. However, these methods remain largely restricted to simple datasets and vision tasks. Their effectiveness on real-world, noisy, and temporally structured biomedical signals, such as ECG, has not been systematically validated.

\begin{figure}[htbp]
  \centering
  \includegraphics[width=0.45\textwidth,height=0.45\textheight]{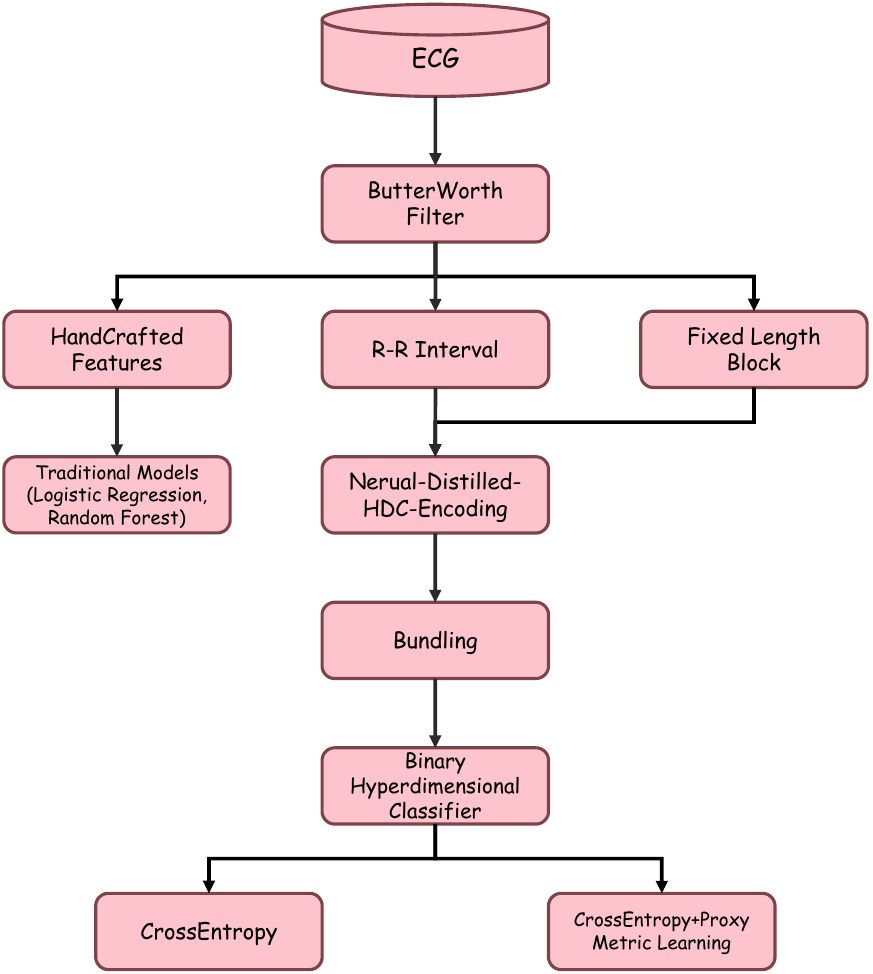}
  \caption{Flowmap}
  \label{fig:flow}
\end{figure}

To bridge this gap, we adopt and extend the \textbf{neural-distilled HDC} paradigm for ECG-based disease detection. Specifically:
\begin{itemize}
\item A lightweight neural encoder is used in place of random item memory, enabling task-adaptive projections of RR-aligned signal blocks;
\item Class hypervectors are optimized using proxy-based metric learning, enhancing inter-class margins beyond conventional averaging;
\end{itemize}

Although we use a softmax classifier during training, this is functionally consistent with cosine-similarity  inference in classical HDC. Previous work~\cite{ma2024hyperdimensional} has shown that softmax outputs over L2-normalized embeddings with temperature scaling approximate cosine similarity rankings, especially when class vectors (proxies) are also normalized. As a result, the network learns to optimize angular margins between hypervectors, aligning with HDC's symbolic inference paradigm.
This hybrid design retains the symbolic structure and computational efficiency of classical HDC, while improving adaptability and discrimination. It also aligns with architectural analogies drawn between HDC and DNNs in~\cite{ma2024hyperdimensional}, and demonstrates that neural-distilled HDC can generalize to real physiological tasks beyond synthetic or vision benchmarks.

\subsection{Flowmap}
We provide an overview of our overall workflow. As illustrated in Fig.~\ref{fig:flow}, there are two main approaches for predicting ECG status. The first approach leverages traditional handcrafted features, such as heart rate, frequency-domain characteristics, and time-domain metrics, which are then used to train a conventional classifier (e.g., logistic regression). The second approach adopts an end-to-end framework that directly utilizes raw ECG signals as input. A preprocessing step is applied to remove baseline drift and high-frequency noise, such as a bandpass Butterworth filter with a passband of 0.5–45 Hz. This method contains two main variants. The first variant segments ECG signals into medically meaningful intervals, such as R-R intervals, and encodes them using HDC techniques. The second, more straightforward variant divides the signals into fixed-length blocks and applies bundling accordingly. In our study, we adopt a neural-distilled HDC encoding strategy, which combines the representational power of neural networks with the efficiency and robustness of HDC. For model training, we compare two loss configurations: standard Cross-Entropy Loss alone versus a combination of Cross-Entropy and Proxy Metric Learning. The implementation details of each component will be discussed in the following sections.

\subsection{Learnable Symbolic Projection for RR Block Embeddings}
To enable symbolic hyperdimensional computing (HDC) while preserving the benefits of gradient-based optimization, we implement a trainable projection module that encodes each RR-interval block into a high-dimensional vector. This module is designed to support both continuous training and discrete inference, aligning with HDC’s symbolic reasoning paradigm.

\subsubsection{Training with Binary-Weighted Projection}
Given an input block \( x \in \mathbb{R}^d \), we apply a linear transformation with learnable weights \( W \in \mathbb{R}^{D \times d} \). To enable binarization while maintaining gradient flow, we adopt a straight-through estimator (STE) strategy. Specifically, we first compute a scaled binary proxy:
\begin{align}
s &= \mathrm{mean}(|W|) \\
\tilde{W} &= s \cdot \mathrm{sign}(W)
\end{align}
The forward pass uses the binarized weights, while gradients are routed through a clipped surrogate:
\begin{equation}
W_{\text{bin}} = \tilde{W}_{\text{stop}} + \left[ \mathrm{clip}(W, -1, 1) - \mathrm{clip}(W, -1, 1)_{\text{stop}} \right]
\end{equation}
where \(\tilde{W}_{\text{stop}} = \mathrm{detach}(\tilde{W})\) and similarly for the clipped term. This construction ensures that only the clipped path contributes to the backward gradient, implementing the STE behavior.

Each block is then encoded as:
\begin{equation}
z = \tanh(W_{\text{bin}} x)
\end{equation}
where the \texttt{tanh} activation serves as a smooth approximation to the \texttt{sign} function. This encourages the latent embedding to converge toward a discrete space while retaining gradient flow.

\begin{figure}[htbp]
  \centering
  \includegraphics[width=\linewidth, height=0.45\textheight, keepaspectratio]{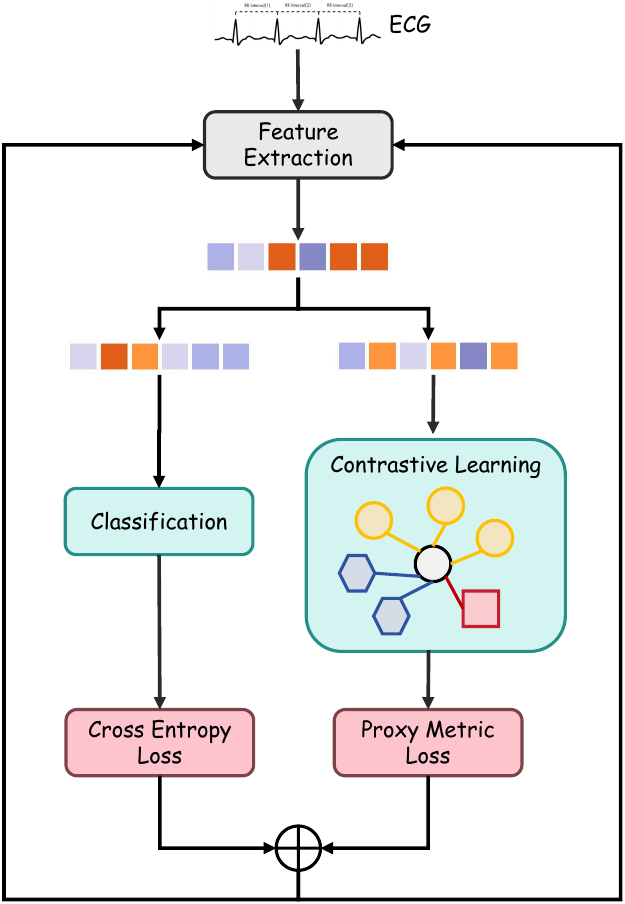}
  \caption{
Neural-symbolic joint training framework with cross-entropy and proxy-based contrastive objectives. The model combines physiological RR block segmentation with learnable symbolic embedding for efficient and interpretable ECG classification.
}
  \label{fig:flow3}
\end{figure}

\subsubsection{Temporal Aggregation}
Given an input tensor \( X \in \mathbb{R}^{B \times T \times L} \), representing a batch of \( B \) samples, each with \( T \) RR blocks of length \( L \), we independently project each block:
\begin{equation}
z_{b,t} = \tanh(W_{\text{bin}} x_{b,t}), \quad z_{b,t} \in \mathbb{R}^D
\end{equation}
and aggregate the resulting high-dimensional vectors across time:
\begin{equation}
Z_b = \sum_{t=1}^{T} z_{b,t}, \quad Z_b \in \mathbb{R}^D
\end{equation}

\subsubsection{Symbolic Inference.}
During inference, we apply a hard discretization to obtain symbolic representations:
\begin{align}
z_{b,t}^{\mathrm{HDC}} &= \mathrm{sign}(z_{b,t})\\
Z_b^{\mathrm{HDC}} &= \sum_{t=1}^{T} z_{b,t}^{\mathrm{HDC}}
\end{align}
resulting in a binary hypervector that supports symbolic comparison. Classification is performed by computing the cosine similarity between \( Z_b^{\mathrm{HDC}} \) and stored class prototypes or associative memories.
\subsubsection{Blockwise Explainability and Temporal Localization}
In addition to producing a global symbolic embedding for classification, our framework inherently supports temporal localization of abnormal patterns through blockwise analysis. Since each RR interval block is individually projected and discretized as \( z_{b,t}^{\mathrm{HDC}} \), we can assess its contribution to the final decision by computing its similarity to class-specific prototype vectors.

\begin{table*}[htbp]
\centering
\caption{Sequence-level classification results on the Apnea-ECG dataset. }
\label{tab:apnea-results}
\resizebox{\linewidth}{!}{
\begin{tabular}{lcccc}
\toprule
\textbf{Model} & \textbf{Accuracy (\%)} & \textbf{Precision (\%)} & \textbf{Recall (\%)} & \textbf{F1-score} \\
\midrule
NeuroHD-RA (Fixed-length Block)         & 46.88 ± 6.35 & 40.26 ± 2.34 & \textbf{78.30 ± 14.92} & 0.524 ± 0.020 \\
Logistic Regression (hand-crafted) & 67.73 ± 1.14 & 57.24 ± 4.38 & 57.66 ± 2.12 & 0.575 ± 0.058 \\
Random Forest (hand-crafted)       & 65.71 ± 2.12 & 58.14 ± 4.12 & 33.01 ± 1.89 & 0.421 ± 0.020 \\
KNN (hand-crafted)                 & 62.28 ± 2.97 & 50.09 ± 6.08 & 48.48 ± 2.12 & 0.493 ± 0.023 \\
NeuroHD-RA (RR Block + CE)              & 61.38 ± 1.64 & 52.75 ± 2.03 & 66.46 ± 6.58 & 0.586 ± 0.019 \\
NeuroHD-RA (RR Block + CE + Proxy)      & \textbf{73.09 ± 0.98} & \textbf{73.68 ± 4.35} & 56.60 ± 3.49 & \textbf{0.626 ± 0.123} \\
ResNet (ECG)                      & 66.03 ± 4.01 & 55.01 ± 1.92 & 59.07 ± 1.48 & 0.571 ± 0.038 \\
AlexNet (ECG)                     & 62.67 ± 1.98 & 50.12 ± 1.16 & 76.12 ± 1.24 & 0.610 ± 0.021 \\
\bottomrule
\end{tabular}}
\end{table*}

Given a class vector \( c \in \mathbb{R}^D \), such as one representing an apneic pattern, the cosine similarity score between each symbolic block embedding and the prototype provides a quantitative estimate of local class affinity:
\begin{equation}
\mathrm{score}_{b,t} = \cos(z_{b,t}^{\mathrm{HDC}}, c)
\end{equation}
These per-block scores can be visualized as a temporal relevance trace over the input sequence, highlighting which RR intervals most strongly align with the abnormal class.

This mechanism enables interpretable predictions and facilitates anomaly detection at high temporal resolution. In clinical applications such as sleep apnea detection, it allows pinpointing specific cardiac segments associated with respiratory disturbances, rather than producing only coarse subject-level predictions.

\subsection{Joint Training with Classification and Contrastive Learning}

To simultaneously achieve accurate label prediction and structured representation learning, we adopt a joint training paradigm that combines classification loss with a contrastive learning (metric learning), as shown in Fig.~\ref{fig:flow3}. This dual-objective strategy enables the model to learn embeddings that are not only discriminative with respect to class boundaries but also semantically organized in the latent space.

Formally, the total training loss is defined as:
\begin{equation}
\mathcal{L}_{\text{total}} = \mathcal{L}_{\text{cls}} + \lambda \mathcal{L}_{\text{contrastive}}
\end{equation}
where $\mathcal{L}_{\text{cls}}$ is the standard cross-entropy loss for supervised classification, $\mathcal{L}_{\text{contrastive}}$ is a metric learning loss that promotes intra-class compactness and inter-class separability, and $\lambda$ is a hyperparameter controlling the contribution of the contrastive term.

To implement the contrastive learning efficiently and scalability, we employ a proxy-based formulation~\cite{movshovitz2017no}, in which each class is represented by a learnable proxy vector in the embedding space. Instead of computing pairwise distances between samples, the loss operates between each sample and the proxies of all classes:
\begin{equation}
\mathcal{L}_{\text{contrastive}} = - \log \frac{\exp(-d(z, p_y))}{\sum_{j=1}^{C} \exp(-d(z, p_j))}
\end{equation}
where $z \in \mathbb{R}^D$ is the embedding of the input sample, $p_y$ is the proxy corresponding to the ground-truth label $y$, $p_j$ is the proxy for class $j$, $C$ is the total number of classes, and $d$ denotes a distance metric (e.g., cosine or Euclidean distance). This formulation provides a more stable and scalable alternative to pairwise or triplet-based contrastive losses.

The joint training framework is especially advantageous in biomedical applications, where data often exhibit high intra-class variability and limited sample sizes. By encouraging a well-structured embedding space, the contrastive learning facilitates better generalization, robustness to distributional shifts, and interpretability. Moreover, the learned representations can be reused for downstream tasks such as disease subtype clustering, anomaly detection, or cross-cohort adaptation without retraining the entire model.

Overall, the synergy between classification supervision and contrastive alignment enables more robust and semantically meaningful feature learning, making the approach well-suited for high-stakes clinical tasks.

\begin{table*}[htbp]
\centering
\caption{Sequence-level binary classification results on the PTB-XL dataset (normal vs apnea). }
\label{tab:apnea-results-binary}
\resizebox{\linewidth}{!}{
\begin{tabular}{lcccc}
\toprule
\textbf{Model} & \textbf{Accuracy (\%)} & \textbf{Precision (\%)} & \textbf{Recall (\%)} & \textbf{F1-score} \\
\midrule
NeuroHD-RA (RR Block + CE + Proxy)      & 70.23 ± 3.27 & 72.92 ± 8.93 & 74.56 ± 8.96 & 0.715 ± 0.072 \\
Logistic Regression (hand-crafted) & 67.58 ± 1.32 & 72.12 ± 1.17 & 64.52 ± 1.97 & 0.681 ± 0.015 \\
Random Forest (hand-crafted)       & 69.25 ± 1.09 & 72.26 ± 1.20 & 69.68 ± 1.20 & 0.702 ± 0.010 \\
KNN (hand-crafted)                 & 65.16 ± 0.76 & 69.29 ± 0.66 & 63.46 ± 1.45 & 0.662 ± 0.009 \\
AlexNet (ECG)                      & 74.36 ± 1.23 & 78.78 ± 1.74 & 75.10 ± 2.53 & 0.764 ± 0.013 \\
ResNet (ECG)                       & \textbf{76.28 ± 1.20} & \textbf{79.45 ± 3.44} & 74.95 ± 3.89 & \textbf{0.775 ± 0.009} \\
\bottomrule
\end{tabular}}
\end{table*}

\section{Experiments And Results}
\subsection{Implementation Details and Evaluation Metrics}

All models are implemented in PyTorch and trained using the Adam optimizer with an initial learning rate of $1 \times 10^{-3}$ and a batch size of 256. The loss balancing coefficient $\lambda$ for the contrastive learning is empirically set to 0.5 based on validation performance. Both the projection weights and the class proxy vectors are optimized jointly in an end-to-end manner. To ensure consistent similarity comparisons, all embeddings and proxies are L2-normalized.

We evaluate the models on two benchmark ECG datasets. The Apnea-ECG dataset comprises long-term single-lead ECG recordings annotated with sleep apnea events. It is specifically designed to support the detection of temporal anomalies indicative of apnea episodes. In contrast, the PTB-XL dataset is a large-scale collection of 12-lead ECG recordings encompassing a broad spectrum of cardiac conditions. These two datasets pose complementary challenges: Apnea-ECG emphasizes fine-grained temporal pattern recognition in a single-channel setting, whereas PTB-XL involves multi-class classification under more diverse cardiac pathologies.


For the Apnea-ECG dataset, we follow the official evaluation protocol and use the predefined test set. The classification task is binary, aiming to detect the presence or absence of sleep apnea. For the PTB-XL dataset, we adopt a \textbf{patient-wise 5-fold cross-validation} strategy to ensure rigorous and unbiased performance evaluation. Specifically, subjects are partitioned into five non-overlapping subsets; in each fold, one subset is held out as the test set, while the remaining four are used for training and validation. This subject-level split prevents any patient overlap between training and testing, thereby avoiding data leakage and better simulating real-world deployment scenarios where generalization to unseen individuals is required.

To reflect the constraints of resource-limited settings such as wearable or ambulatory devices, we restrict our analysis to a single lead, Lead I. This configuration represents a minimal-sensor deployment scenario and poses a more challenging yet practically relevant evaluation setting.

As baselines, we include classical machine learning methods (logistic regression, random forest), as well as convolutional neural networks (CNNs). Specifically, we use a 1D variant of ResNet18 adapted for single-channel ECG inputs. The model architecture and kernel sizes are adjusted accordingly to match the temporal resolution of 1D signals. All CNN baselines are trained using standard cross-entropy loss in raw ECG segments preprocessed with a Butterworth bandpass filter (0.5 to 45 Hz).
We report average performance across all folds using standard classification metrics, including accuracy, precision, recall, and F1 score. To assess the practical efficiency of the models, we additionally report the average inference time per sample (in milliseconds), providing insights into their suitability for real-time or edge-based deployment.

All the reported results for our proposed method are based on the high-dimensional configuration (\( D = 10{,}000 \)), which strikes a balance between accuracy and symbolic representational capacity. This dimensionality was chosen to highlight the full expressive power of our neural-distilled HDC framework while still maintaining practical inference efficiency.

\subsection{Results on Apnea-ECG}

We evaluate the proposed framework on the widely used Apnea-ECG benchmark dataset~\cite{penzel2000apnea}, which is designed for detecting sleep apnea events from single-lead ECG signals. Table~\ref{tab:apnea-results} summarizes the sequence-level classification performance of our approach and several baselines, including traditional machine learning models using handcrafted features and neural network-based models trained on raw ECG signals. All results are reported as the mean ± standard deviation across three random seeds.

Our RR block-based method with contrastive loss and proxy-based supervision (\textbf{RR Block + CE + Proxy}) achieves the best overall performance, reaching an accuracy of \textbf{73.09\%} and an F1 score of \textbf{ 0.626}, outperforming both neural baselines such as ResNet and AlexNet, and conventional models such as SVM, logistic regression, and random forest. In particular, this model provides a good balance between precision and recall, highlighting the effectiveness of our rhythm-aware representation and the integration of proxy supervision in enhancing discriminative ability.

In contrast, the fixed-length block variant, although it achieved a high recall (78.30\%) suffers from very low precision and overall accuracy, suggesting that naive segmentation without aligning to cardiac cycles leads to substantial information loss. This underscores the necessity of building rhythm-aligned features for the detection of apnea.

Traditional models using hand-made features (e.g., RR statistics, morphological features) perform moderately well, with logistic regression and SVM reaching F1 scores around 0.57. Although these models are interpretable and computationally efficient, they lack the capacity to model complex temporal dynamics compared to our proxy-enhanced HDC pipeline. In particular, the random forest baseline shows relatively high precision, but suffers from a significant drop in recall, likely due to overfitting to limited local patterns.

Compared to 1D-CNN-based architectures, our RR-based HDC framework shows competitive or superior performance. For example, although AlexNet achieves a comparable recall (76.12\%), its overall F1 score remains lower (0.610), likely due to its inability to explicitly model the temporal structure between beats. Together, these results demonstrate the advantage of combining rhythm-aware representation, high-dimensional encoding, and proxy-based supervision.


\subsection{Results on PTB-XL}
We further evaluate our framework on the PTB-XL dataset under a binary classification setting (normal vs. abnormal). Results are also reported as mean ± standard deviation over multiple runs.
As summarized in Table~\ref{tab:apnea-results-binary}, our RR-block-based HDC model with cross-entropy and proxy-based supervision achieves robust performance, reaching an F1-score of \textbf{0.715}. Although deep learning baselines such as ResNet and AlexNet outperform our model slightly in terms of accuracy and F1-score, they require significantly more computation and model complexity. In contrast, our method strikes a strong balance between performance, interpretability, and computational cost.

In addition, traditional classifiers trained on hand-crafted features, such as logistic regression, random forest, and k-nearest neighbors, exhibit competitive precision (e.g., 72.26\%) but fall short in terms of F1-score due to weaker recall and generalization. This further highlights the strength of rhythm-aware symbolic representation combined with discriminative learning.

\subsection{Inference Efficiency}

\begin{table}[htbp]
\centering
\caption{Inference Time Comparison}
\label{tab:cpu_time}
\begin{tabular}{lc}
\toprule
\textbf{Model} & \textbf{Inference Time (ms)} \\
\midrule
NeuroHD-RA (Binary) & $21.54 \pm 0.45$ \\
AlexNet1D (Float32)               & $28.50 \pm 0.90$ \\
ResNet18 1D (Float32)             & $33.85 \pm 1.10$ \\
\bottomrule
\end{tabular}
\end{table}

To assess the deployment feasibility of our proposed model, we systematically evaluated its inference efficiency and memory footprint on a standard CPU platform. As shown in Table~\ref{tab:cpu_time}, our neural-distilled HDC model with high-dimensional binary encoding (\( D = 10000 \)) achieves a per-sample inference time of \textbf{21.54~±~0.45~ms}, outperforming conventional deep learning baselines such as AlexNet1D (\textbf{28.50~±~0.90~ms}) and ResNet18 1D (\textbf{33.85~±~1.10~ms}).

\begin{table}[htbp]
\centering
\caption{Model Compression Rate Comparison}
\label{tab:compression_rate}
\resizebox{\linewidth}{!}{%
\begin{tabular}{lcc}
\toprule
\textbf{Model} & \textbf{Param Size (KB)} & \textbf{Compression Rate} \\
\midrule
NeuroHD-RA (Binary)             & 124.5   & $1\times$ \\
AlexNet1D (Float32)    & 3,080   & $24.7\times$ \\
ResNet18 1D (Float32)  & 12,560  & $100.8\times$ \\
\bottomrule
\end{tabular}
}
\end{table}

The complementary results in Table~\ref{tab:compression_rate} further reveal a dramatic reduction in the size of parameters. Our model requires only \textbf{124.5~KB}, achieving a compression rate of up to \textbf{100.8$\times$} compared to ResNet18-1D and \textbf{24.7$\times$} compared to AlexNet-1D. This significant reduction in model size is made possible by our symbolic, binary-weighted architecture, which eliminates convolutional layers and replaces them with lightweight feedforward computations.

The efficiency stems from the symbolic, fully feedforward nature of our HDC-based inference pipeline, which eliminates the need for expensive convolutional operations and deep sequential layers. Furthermore, the use of binary weights drastically reduces computational overhead and memory footprint, making  the model particularly suitable for edge deployment scenarios such as wearable health monitors and real-time biomedical signal processing on resource-constrained IoT devices. In conjunction with the competitive classification performance demonstrated earlier, our approach offers a compelling balance of accuracy, interpretability, and hardware efficiency for scalable biomedical applications.

\subsection{Visualization}
\begin{figure}[htbp]
  \centering
  \includegraphics[width=\linewidth, height=0.45\textheight, keepaspectratio]{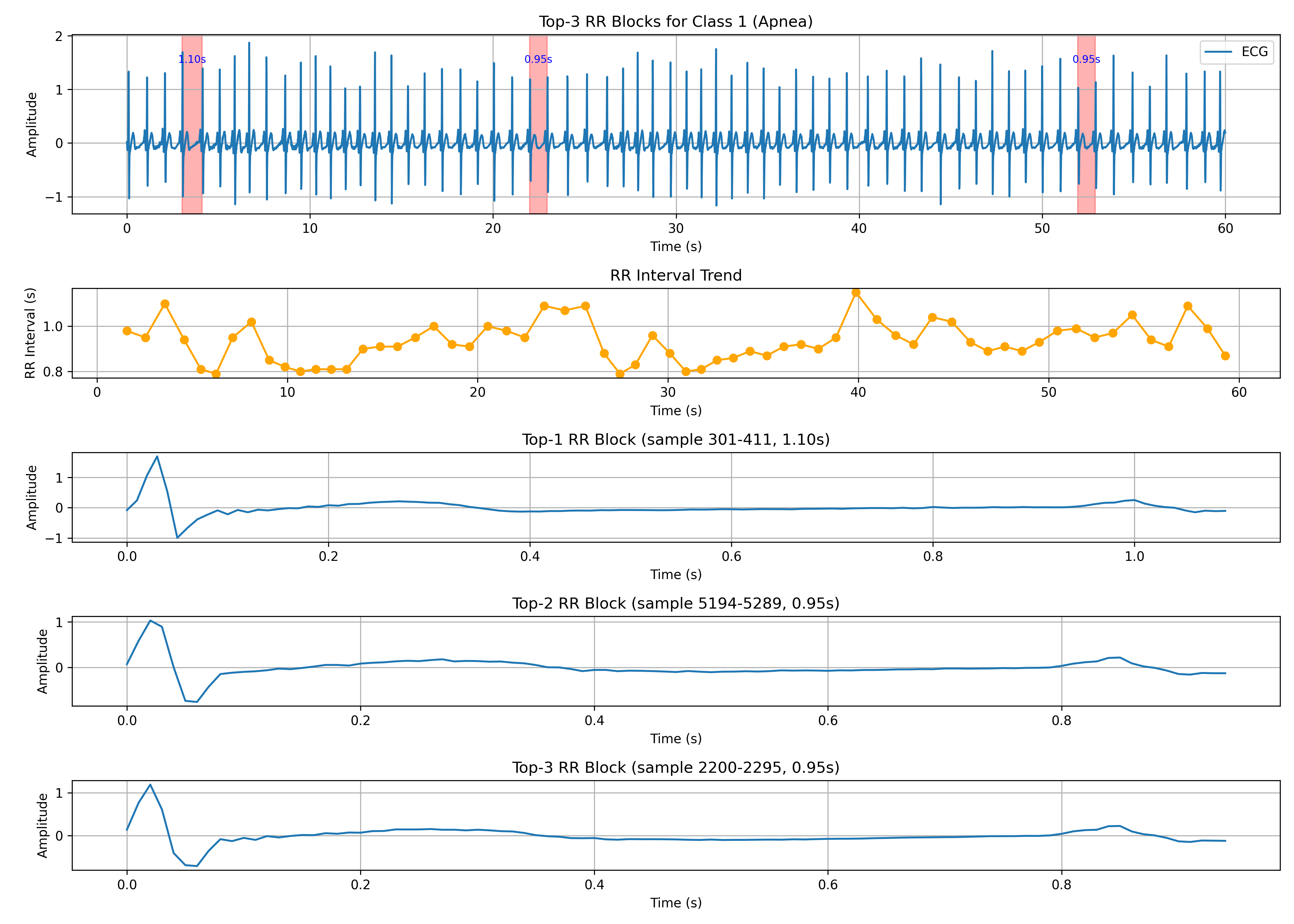}
  \caption{Interpretable Visualization of RR-Interval Blocks}
  \label{fig:visualization}
\end{figure}

To enhance interpretability and gain insights into the decision-making process of our model, we visualize the RR-interval blocks extracted from raw ECG segments alongside the regions most influential to the model's prediction. As illustrated in Fig.~\ref{fig:visualization}, each segment is partitioned according to detected R peaks to form RR blocks, which are later encoded in high-dimensional embeddings. We then highlight the top-$k$ blocks that exhibit the highest class-specific contribution scores, as determined by the learned projection and classification modules.

This composite visualization offers a dual perspective: It reveals the physiological structure of the RR intervals while simultaneously exposing the discriminative focus of the model. This alignment between physiological signals and model attention provides an interpretable basis for classification outcomes, particularly critical in clinical contexts where explainability is paramount. In particular, highlighted blocks often correspond to irregular or elongated RR intervals, which are consistent with the pathological patterns associated with sleep apnea. This validates the effectiveness of our learnable symbolic projection mechanism in capturing semantically meaningful and diagnostically relevant features from compact RR representations.
\section{Discussion and Limitation}

Our study demonstrates that symbolic hyperdimensional computing (HDC), when guided by physiological priors and enhanced through learnable projections, can serve as an efficient and interpretable alternative to deep learning models for the detection of ECG abnormalities. By aligning HDC representations with R-R interval blocks, and incorporating proxy-based supervision into a lightweight symbolic architecture, we achieve competitive performance while preserving symbolic reasoning capabilities.

One key advantage of our approach lies in its interpretability. Unlike deep neural networks, which often require post hoc explainability tools, our model enables block-level semantic tracing via projection similarity. This allows direct visualization and localization of R-R interval blocks that contribute most to classification decisions, offering a clinically meaningful interpretation pathway. Moreover, the use of binary-weighted projections and bundling operations supports bit-level composability, a desirable property in logic-driven and hardware-constrained settings.

Another strength is the efficiency of our inference pipeline. The proposed model maintains a small footprint, avoids convolution or recurrence, and exhibits substantially faster inference compared to CNN-based models. This makes it especially suitable for wearable or resource-limited platforms where energy efficiency and real-time processing are critical.

However, several limitations remain. First, although our method achieves strong performance compared to traditional models and shallow deep learning methods, it still lags behind large-scale deep models in absolute classification accuracy. This reflects the inherent trade-off between model complexity and symbolic fidelity. Notably, our method prioritizes maintaining the symbolic and resource-efficient nature of HDC, rather than purely optimizing for classification performance. Second, our current implementation does not incorporate temporal attention or sequence-level dynamics beyond bundling, which may limit its ability to capture long-term patterns. Although adding such mechanisms (e.g., attention, transformers) could improve accuracy, it may undermine the symbolic nature of the model and increase computational cost.


\bibliographystyle{IEEEtran}
\bibliography{ref}
\end{document}